\documentclass{ws-ijmpd}
\usepackage[super,compress]{cite}

\usepackage{graphicx}

\def\laq{~\raise 0.4ex\hbox{$<$}\kern -0.8em\lower 0.62
ex\hbox{$\sim$}~}
\def\gaq{~\raise 0.4ex\hbox{$>$}\kern -0.7em\lower 0.62
ex\hbox{$\sim$}~}

\def\beq{\begin{equation}}
\def\eeq{\end{equation}}
\def\bea{\begin{eqnarray}}
\def\eea{\end{eqnarray}}

\def \wh {\widehat}
\def \ra {\rightarrow}

\begin{document}

\markboth{M. Gasperini}
{Do alien particles exist, and can they be detected?}

%
\catchline{}{}{}{}{}

%

\title{
DO ALIEN PARTICLES EXIST, AND CAN THEY BE DETECTED?\footnote{ Essay written for the Gravity Research Foundation, {\em 2016 
Awards for Essays on Gravitation.} 
To appear in the October 2016  Special Issue of International Journal of Modern Physics D.}}

\author{M. GASPERINI}

\address{Dipartimento di Fisica,
Universit\`a di Bari,\\ Via G. Amendola 173, 70126 Bari, Italy,\\
and Istituto Nazionale di Fisica Nucleare, Sezione di Bari,\\ Via E. Orabona 4, 70125 Bari, Italy\\
gasperini@ba.infn.it}

\maketitle


\begin{abstract}
We may call ``alien particles" those particles belonging to the matter/field content of a $d$-dimensional brane other than the $3$-brane (or stack of branes) sweeping the space-time in which we live. They can appear in our space-time at the regions of intersection between our and their brane. They can be identified (or not) as  alien matter depending on their properties, on the physical laws governing their evolution in the ``homeland" brane, and on the details of our detection techniques.
\end{abstract}


\ccode{{\em Preprint number}: BA-TH/702-16}



\vspace{1cm}

Modern physical theories -- in particular, higher-dimensional unified models of gravity and of the other fundamental interactions -- have accustomed us to the idea that the space-time in which we live can be appropriately represented as the four-dimensional hypersurface spanned by the evolution of a $3$-brane, embedded in an external ``bulk" manifold. We may recall, in this respect, the so-called brane-world scenario, and its many physical/cosmological applications (see e.g. Ref. 1 for an updated review).

The bulk, however, might also contain many other branes similar to ours. The space-times spanned by the various branes might have reciprocal (and possible multiple) intersections. The intersection of (wrapping, Dirichlet) branes is indeed at the ground of possible string-theory explanations of basic Standard Model properties, such as the chirality of the fermion spectrum and the observed number of quark/lepton generations \cite{Ibanez}. But there are also cosmological and inflationary consequences of brane intersections \cite{Bellido,Gomez}.

Quite independently of the possible motivations/applications mentioned above, another important aspect of brane intersections, in my opinion, is the following. 
Even if the matter/field content of a given brane  is rigidly localized (excluding gravity \cite{1}) on the associated world-volume, a direct interaction among the matter components of different branes (or stacks of branes) turns out to be possible, in principle, at the intersection regions, where the world-lines of particles belonging to different branes can mix and intersect. Those regions may thus behave as open windows on the extra dimensions. 

Let us consider, in such a context, the high-energy collision of two particles $p_1$ and $p_2$ belonging to our brane $D_3$, and suppose that the collision point lies within the region of spatial intersection with another brane $\wh D_3$. In that case the collision can possibly produce particles ($\wh p_3$, $\wh p_4$, \dots)
which are living on the brane  $\wh D_3$ (let us call them ``alien" particles), but which nevertheless propagate, are visible, and can possibly be detected also inside our brane -- at least until they keep inside the region of spatial intersection\footnote{Note that the shape and the size of such a region can be time-dependent, depending on the state of relative motion through the bulk of the two branes, and/or on the geometric (and topological) properties of the two space-time manifolds (see Fig. 1).}.

If the two branes $D_3$ and $\wh D_3$ are identical and, in particular, characterized by the same particle content, then no anomalous effect associated to the process $p_1+p_2 \ra \wh p_3 + \wh p_4 + \cdots$ can obviously be observed. The particle content of the two branes, however, could be different; in addition, the physical properties of the so-called alien particles could differ, in principle, from the properties of the particles belonging to our brane.

\begin{figure}[t]
\centering
\includegraphics[height=6.5 cm]{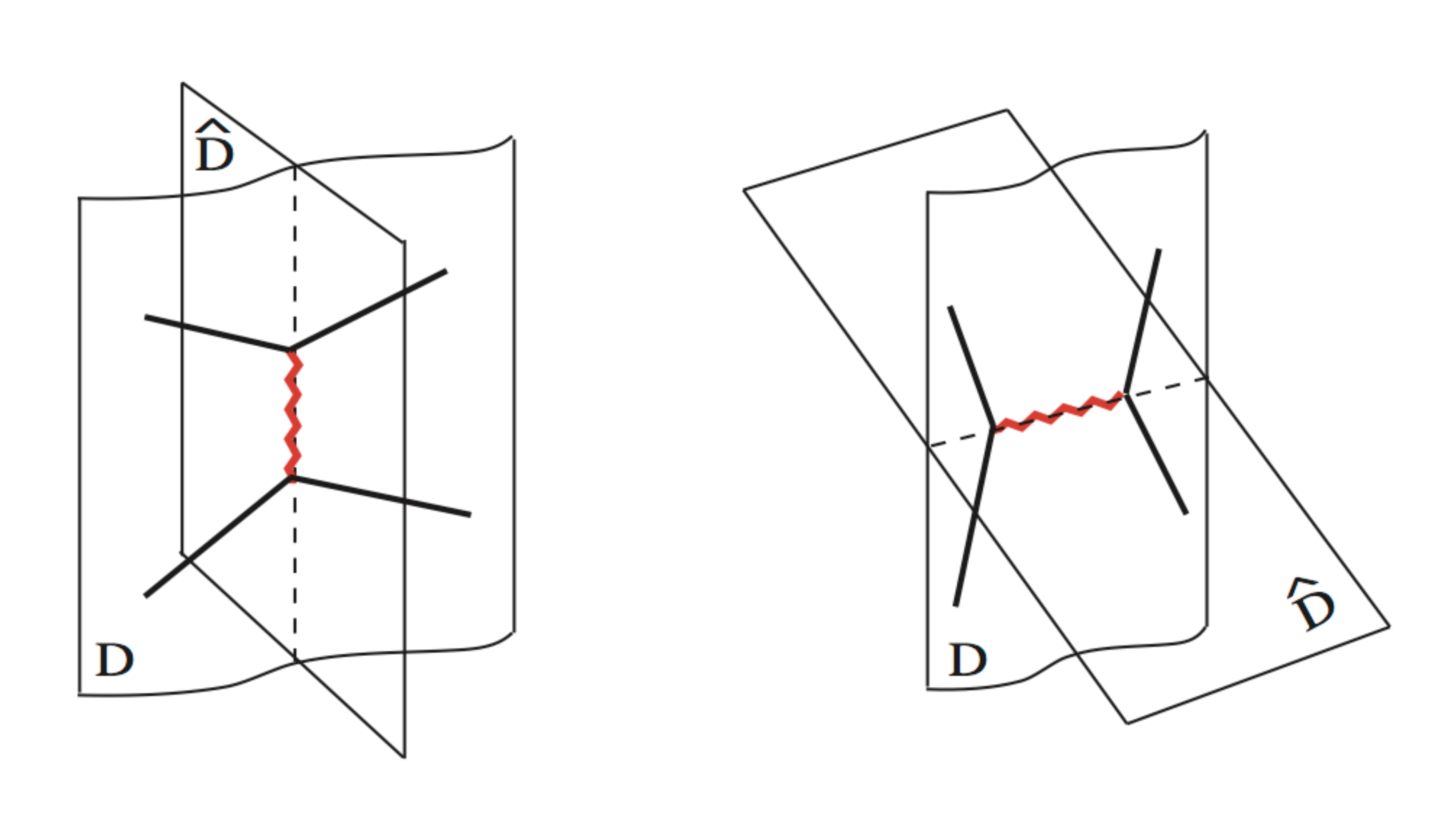}
\caption{The world-sheets of two branes $D$ and $\wh D$ with timelike (left) and spacelike (right) intersections. We have also plotted two possible quantum processes localized in our brane $D$, and described by Feynman diagrams with external legs in $D$ and internal legs in the region of intersection with $\wh D$. The particle propagating in the intersection region is virtual in the right picture, and real in the left one.}
\label{Fig1}
\end{figure}

For a simple illustration of this possibility we may refer those models allowing (and predicting) the time and space variation of the so-called fundamental constants\footnote{There are many possible examples of models of this type, triggered by an idea suggested long ago by Dirac \cite{2}, first implemented (for the variation of the Newton constant) by scalar-tensor theories of gravity \cite{3} and later extended to the variation of the electromagnetic coupling \cite{4} and of other physical parameters (see e.g. Ref. 8).  See in particular Ref. 9 for a review of the theoretical/phenomenological aspects of this scenario and a discussion of its possible experimental verifications. But see also the so-called ``self-reproducing" inflationary models \cite{7}, and the associated ``multiverse" scenario \cite{Don} (strongly motivated by string theory), where the space-time may have different four-dimensional domains with different physical constants and different low-energy phenomenology.}. According to such models a theory of fundamental interactions has in general many possible ground states (or vacua), with possibly different realizations (and different physical constants) in different bulk domains: hence, the alien particles living on $\wh D_3$ might obey physical laws characterized by a set of fundamental parameters different from those typical of our brane $D_3$.

We shall concentrate, in this Essay, on the particularly interesting case in which the physics and the particle content of the two branes $D_3$ and $\wh D_3$ are the same except for the fact that, on the $\wh D_3$ brane, light propagates in vacuum with a constant velocity $\wh c \not=c$ (let us suppose, in particular, 
$\wh c >c$). Hence, alien particles satisfy the usual relativistic laws with $c$ replaced by $\wh c$. This, of course, can break the global Lorentz symmetry of the higher-dimensional bulk manifold, but has no effect on the (separate) validity of the relativistic Lorentz invariance in the four-dimensional space-times spanned by the two branes\footnote{The possibility of a speed of light different from $c$ in localized (microscopic) space-times domains was first considered, to the best of my knowledge, in Refs. 12 and 13. More recently, the possible variation of the speed of light in vacuum has been proposed and discussed also at a macroscopic and cosmological level \cite{9,10,11} (see Ref. 17 for a detailed review).}.

The obvious question which arises, in this case, is the following: can we discover the anomalous behavior of this type of particles, and unambiguously classify them as alien particles? The answer crucially depends on our detection techniques. 

If the life of an alien particle inside our brane is long enough to allow for direct velocity measurements (of the particle itself, or of its alien decay products) then, of course, the anomaly might explicitly appear whenever we could find for the velocity $v$ a result in the range $c < v < \wh c$. Even if $v<c$, however, the anomaly might appear when, measuring also the relativistic momentum $\vec p$ and energy $E$ of the particle, we could find that 
\beq
\vec p \not= {E \vec v \over c^2 }
\eeq
(since, for the alien particle, $\vec p =E \vec v /\wh c^2$). 

Similarly, if the type of particle (and its mass) is known, and if we could be able to measure both $\vec v$ and $\vec p$, the anomaly might appear whenever 
\beq
\vec p (v) \not= {m\vec v \over(1- v^2/c^2)^{1/2}}
\eeq
(since, for the alien particle, the effective Lorentz factor depends on the ratio $v^2/\wh c^2$).

The situation is different, however, in the more realistic case of very short-lived particles\footnote{This case is more realistic because alien particles can be present in our brane $D_3$ only until they are inside the region of intersection with their brane, a region expected to have, in general, very limited spatial (and temporal) extension.}  (like, for instance, a typical hadronic resonance).  In that case there is no hope (at present) to measure its velocity, and the only kinematical variable that we can associate to the particle is the so-called ``invariant mass" parameter $M$, obtained from the total energy $E_{CM}$ of its decay products in the center of mass system by applying the standard relation $M= E_{CM}/c^2$.

This relation is incorrect, of course, for alien particles, for which $\wh M= E_{CM}/\wh c^2$. This leads to overestimate the real mass of the particle by the factor
\beq
{M/ \wh M}= \left(\wh c/ c\right)^2. 
\eeq
This discrepancy could only appear by comparing the kinematic parameter $M$ with complementary experimental information on the actual mass value, obtained for instance from measures concerning the particle gravitational interactions. Barring this, as well other, presently unlikely, possibilities, we can ask: are there particle states which might have been misinterpreted, up to now, in this way? 

The question may be legitimate, in principle, for those states which mainly differ only for their masses and have very similar quantum numbers. Are (some of) these states alien particles? Is the width of such states (i.e., the associated mean life)  determined by the size (or the ``thickness') of the intersection region between our brane and the alien brane?  Is our Universe, at the subnuclear level, a dense and tangled network of intertwining brane space-times, continuously overlapping, splitting up, and crossing among each other? 

Maybe the example considered here -- namely, alien branes with different values of $c$ -- is not the most appropriate or the most realistic case to consider. Maybe there are more fundamental parameters varying through different domains of the bulk space-time (let us recall, in this respect, the possible relation between the observed values of the standard Yukawa couplings and the intersection area of different branes \cite{Ibanez1}). Nevertheless, I think there is a possible lesson that we could learn from the above discussion.

In the occurrence of unexpected results in high-energy experiments (like, for instance, the $750$ GeV diphoton excess\cite{13,13a} recently reported at LHC), we should look for possible interpretations not only as signals of new particles not included in the standard phenomenological scenario, but also as signals produced by standard particles crossing our space-time but coming from an ``outside" brane, where physics is the same but (some of) the fundamental physical parameters are appreciably different. 

Remarkably, this approach could also provide a way -- probably the only way -- to test the so-called (and so far experimentally elusive) multiverse scenario \cite{Don}. 
In fact, the existence of different physical domains in very distant, disconnected regions of space-time is likely not directly testable, as stressed in Ref. 11. The local intersections of different branes, on the contrary, could bring the ``alien" space-time domains (i.e., the different ground state regimes) directly inside our world and our accessible physical experience. 

\section*{Acknowledgments}

It is a pleasure to thank Donato Creanza and Salvatore My for useful information on current detection techniques of high-energy particle physics. This work is supported in part by MIUR, under grant no. 2012CPPYP7 (PRIN 2012), and by INFN,  under the program TAsP ({Theoretical Astroparticle Physics}).


\end{document}